\documentclass[epsf]{aastex}

\include{epsf}
\usepackage{emulateapj5}

\shorttitle{Globular Clusters in NGC 1399}
\shortauthors{Forbes, Beasley, Brodie \& Kissler-Patig}

\def\etal{{\it et al. }}

\begin{document}


\title{Age Estimates for Globular Clusters in NGC 1399}


\author{Duncan A. Forbes and Michael A. Beasley}
\affil{Astrophysics \& Supercomputing, Swinburne University,
  Hawthorn, VIC 3122, Australia}
\email{dforbes, mbeasley@swin.edu.au}

\author{Jean P. Brodie}
\affil{Lick Observatory, University of California,
 Santa Cruz, CA 95064, USA}
\email{brodie, soeren@ucolick.org}

\author{Markus Kissler-Patig}
\affil{ESO, Karl-Schwarzschild-Str. 2, 85748 Garching, Germany}
\email{mkissler@eso.org}



\begin{abstract}
We present high signal-to-noise Keck spectra for 10 globular
clusters associated with the giant Fornax elliptical NGC~1399,
and compare measured line indices with current stellar population
models. 
Our data convincingly demonstrate, for the first time, 
that at least some of the
clusters in a giant elliptical galaxy 
have super-solar abundance ratios, similar to the host galaxy.
From H$\beta$ line-strengths the majority of
clusters have ages of $\sim$11 Gyrs (within 2$\sigma$), which is
similar to the luminosity-weighted stellar age of 
NGC~1399. 
Two of the clusters 
(which also reveal enhanced abundance ratios) 
show significantly higher H$\beta$ values than the
others. It remains unclear whether this is due to 
young ($\sim$ 2 Gyr) ages, or extremely old ($>$ 15
Gyr) ages with a warm blue horizontal branch. However a conflict with
current cosmological parameters is avoided if the young age is
favored.  
Either alternative indicates a complicated age distribution among
the metal-rich clusters and sets interesting 
constraints on their chemical enrichment
at late epochs. 

\end{abstract}


\section{Introduction}\label{sec_intro}

The metal-rich and metal-poor Globular Clusters (GCs) in our
Galaxy, are thought to be both very old, i.e. $\sim$13 Gyrs (e.g.
Carretta \etal 2000) and enhanced in $\alpha$ elements, i.e. 
[$\alpha$/Fe] $\sim$ +0.3 (Carney 1996). However the amount, and even the
existence of a relative age difference between these two
subpopulations is a matter of current debate. 


Giant ellipticals also reveal two subpopulations of GCs (e.g. 
Larsen \etal 2001), 
with mean metallicities similar to those seen in spirals
(Forbes, Brodie \& Larsen 2001). Very little is currently 
known about the ages and abundance ratios of such GCs. 
The spectroscopic (Kissler-Patig \etal 1998; Cohen
\etal 1998; Beasley \etal 2000) and recent photometric (Puzia \etal
1999; Kundu \etal 1999) studies suggest that both
subpopulations are very old but the red (metal-rich) GCs could be
up to as much as 5 Gyrs younger than the blue (metal-poor) ones.
Abundance ratios for GCs in the giant ellipticals studied to date are
consistent with Galactic GCs but individual errors are large. 
In the case of the merger remnant NGC 7252, Maraston \etal (2001)
found GCs that had both young ($\sim$ 1 Gyr) ages and {\it solar}
abundance ratios. 
Accurate measurements of individual GC ages and abundance ratios 
in elliptical
galaxies will provide key constraints on the possible GC formation
mechanisms (Ashman \& Zepf 1992; Forbes, Brodie \&
Grillmair 1997).


In our previous spectroscopic study of NGC 1399 we
obtained spectra of 18 GCs with the LRIS spectrograph on the 
Keck telescope (Kissler-Patig \etal 1998). 
These spectra indicated that the two subpopulations were old and
coeval within the errors (i.e. $\pm$ 0.5~\AA~ in H$\beta$ Lick line 
index). Such errors were too large to clearly
distinguish a relative 
age difference between the two subpopulations but 
they were generally consistent with old ages similar to 
those found in the Milky Way GC system. We also identified two
metal-rich GCs with potentially super-solar abundance ratios. 

Here we present higher S/N spectra (i.e. errors on H$\beta$ are
about $\pm$ 0.25~\AA~) for 10 GCs. These data are perhaps
the highest quality spectra published for GCs beyond the Local
Group and are comparable to the best spectra in the literature 
for M31 (Huchra \etal 1996) and Milky Way GCs (Trager \etal 2000). 
Errors in spectral line indices 
have a huge impact on our ability to estimate GC
ages. For our previous dataset these errors translated into
a relative age uncertainty of about $\pm$ 10 Gyrs for a 15 Gyr old GC. The
data presented here have a corresponding error of $\pm$ 5
Gyrs. 
This allows us to better estimate the relative age of individual
GCs and in particular identify GCs with young inferred ages in a giant
elliptical galaxy for the first time. 
Here we confirm that the
metal-poor GCs have [Fe/H] $\sim$ --1.5 and are very old. 
We also identify two metal-rich GCs 
with super-solar abundance ratios (i.e. [$\alpha$/Fe] $>$ 0) -- 
such ratios are generally unexpected in late epoch mergers of
spiral disks (Goudfrooij \etal 2001).

\section{Observations and Data Reduction}

Candidate GCs were selected from the list of
Grillmair (1992) and observed on 1997 Sept. 30th 
and Oct. 1st using the LRIS spectrograph (Oke \etal 1995) on the 
Keck I telescope. The observational
setup was the same as used by Kissler-Patig \etal (1998) except
integrations were longer at 13,200 sec. The 600~l/mm grating gave a 
resolution of 5.6\AA~. 

Data reduction was carried out using the {\tt REDUX} software
package developed by A. Phillips. Using a series of scripts this
package subtracts the bias, flatfields the data, removes the
$x-$ and $y-$distortions, and produces optimal 
sky subtracted 1-D spectra. Comparison lamp spectra of Hg, Ar, Ne
and Kr were used for wavelength calibration. Spectra from the
different nights were combined. Flux calibration was
provided by the flux standard BD284211 observed on the first
night. To correct the GCs onto the Lick/IDS system, 
we convolved our spectra with a wavelength-dependent
Gaussian kernel and then applied
small offsets obtained from observations of several Lick
standard stars (see Beasley \etal in prep.).

Lick indices (Trager \etal 2000) 
were measured from our flux-calibrated spectra. 
Due to the variable nature
of the wavelength ranges in multi-slit spectra, the same set
of indices were not measured for all spectra.
Uncertainties in the indices were derived from the 
photon noise in the unfluxed spectra.
We have obtained spectra with S/N = 30--45 \AA$^{-1}$, 
giving errors in the H$\beta$ index of 0.34--0.22 \AA.

Of the 17 usable spectra we confirm that
11 are {\it bona fide} GCs. 
We found objects \#43 and \#164 (IDs from Grillmair 1992) 
to be Galactic stars. 
Background galaxies (and their redshifts) are \#40 (z$\sim$0.11),
\#163 (z$\sim$0.07), \#167 (z$\sim$0.14) and \#169 (z$\sim$0.13).
Our sample of GCs have
an average galactocentric distance of 20 kpc and cover the
observed range of C--T$_1$ colors for the GC system (Ostrov,
Geisler \& Forte 1993). 
Velocities have been measured from the spectra via cross-correlation
with high S/N spectra of two 
M31 GCs (158-213; $v_{\rm helio}$ = --180 km/s 
and 225-280; $v_{\rm helio}$ = --164 km/s).
The 11 GCs have a mean velocity of 1551 $\pm$ 74 km/s and
velocity dispersion of 246 $\pm$ 57 km/s. NGC 1399 itself has a
velocity of 1447 $\pm$ 12 km/s. 
One GC, \#41 with velocity 1619$\pm$68 km/s, 
has been excluded from our 
line-strength analysis due to suspect sky-subtraction.

\section{Ages and Abundances}

To investigate the properties of our GC sample, 
we primarily 
use the stellar population models of Maraston \& Thomas (2000)
which predict line-strength indices
using the Lick/IDS-based fitting functions of Worthey (1994). 
It is important to check that we have adequately corrected
the data onto the Lick system. To this end, we have compared 
index-index plots of our data with the models, and generally
find good agreement. 
The strongest (and best measured)
features in the GC spectra, which we use in this
study, are the primarily metallicity-sensitive 
Mg$_2$, Mg~$b$ and $\langle$Fe$\rangle$ indices 
(the mean of the Fe5270 and Fe5335), and the more age-sensitive
H$\beta$ and H$\gamma_{\rm A}$ (the broader of the two
H$\gamma$ indices defined by Worthey \& Ottaviani 1997).
Our line index measurements for the 10 GCs in NGC 1399 are
listed in Table 1. We also include T$_1$ magnitudes 
(similar to Johnson R) and C--T$_1$ colors 
from Geisler, Forte \& Dirsch (in prep.).

In Fig.~1 we 
plot the magnesium and iron indices of the GCs, 
and for the 
central line-strength of NGC~1399 itself (taken from Kuntschner 2000).
At low metallicities, the GCs follow the models
reasonably well.
However, at higher metallicities, the NGC~1399 GCs 
deviate significantly from the grids.
The metal-rich GCs seemingly show an enhancement of 
magnesium with respect to iron, $and$ an enhancement of
Mg $b$ with respect to Mg$_2$. 

This behavior is also exhibited by NGC~1399
itself, which lies to the right of the population models.
The fact that these magnesium lines do not vary in the same
fashion does not indicate that we are unable 
to measure these indices. Rather, these indices (i.e. the
bandpasses) have different
contributions from elements other than magnesium (e.g. Tripicco
\& Bell 1995; Trager \etal 2000).
Since the models of Maraston \& Thomas (2000) use
scaled-solar isochrones, we conclude that this offset arises
because the metal-rich GCs in our sample have non-solar abundance ratios, 
i.e. [Mg/Fe] $>$ 0. This result is consistent with high-resolution
spectroscopy of bright giants in Galactic GCs, which typically 
exhibit [$\alpha$/Fe] $\sim$ + 0.3 (Carney 1996).
Due to the small difference between 
solar and $\alpha$-enhanced isochrones at low metallicities 
(e.g. Salaris \& Weiss 1998) it is 
possibile that the metal-poor GCs also possess super-solar abundance
ratios.

\vbox{
\begin{center}
\leavevmode
\hbox{%
\epsfxsize=9.2cm
\epsffile{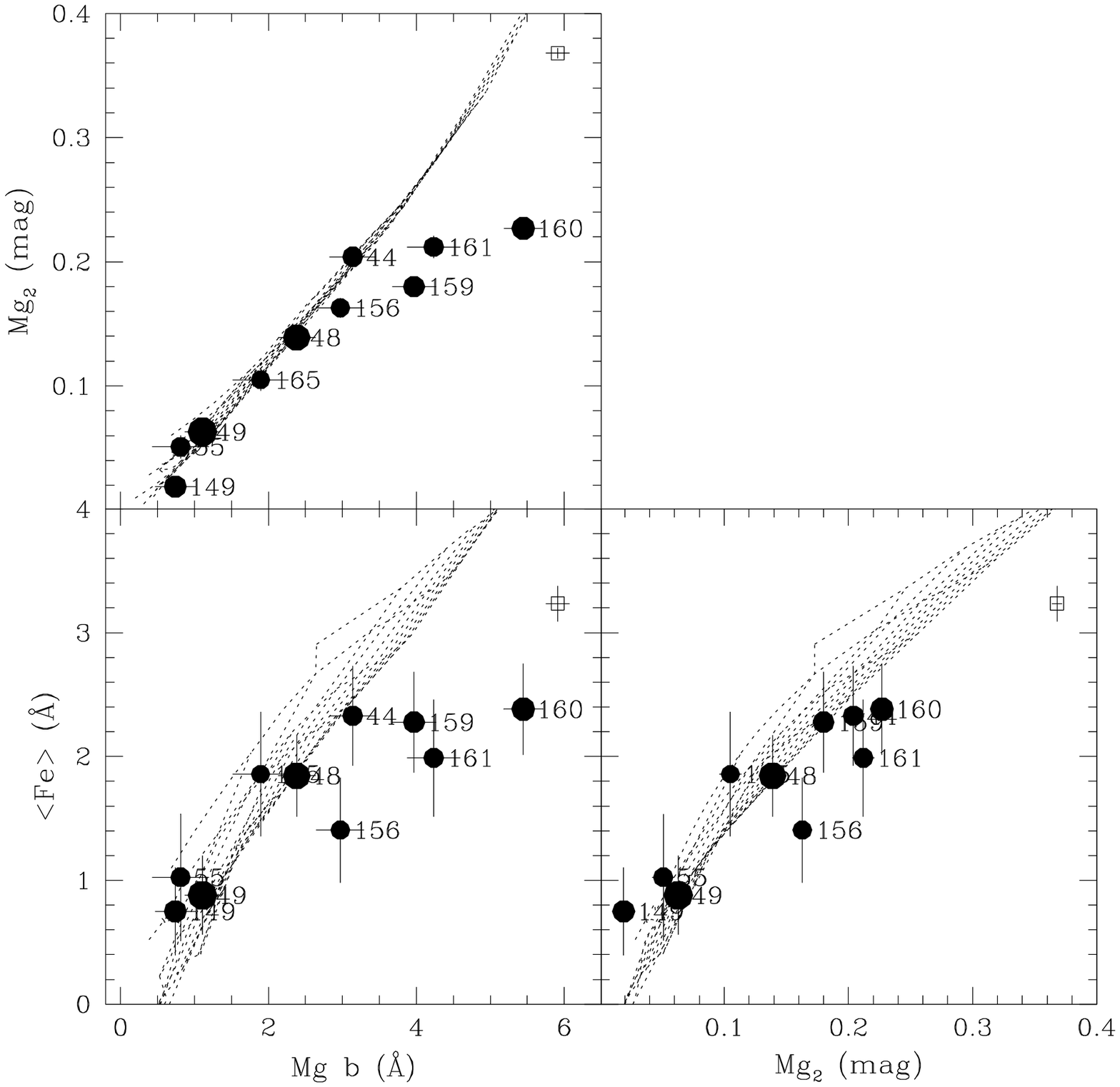}}
\figcaption{\small
Comparison of primarily metallicity-sensitive Lick/IDS indices.
Filled, numbered circles with error bars are the NGC~1399 GCs. 
Symbol size is roughly proportional to the S/N of the
spectra. The open square indicates the position of NGC~1399
(Kuntschner 2000). Over-plotted are the stellar
population models of Maraston \& Thomas (2000) for ages
between 1 and 15 Gyrs (left to right), and --2.25 $<$ [Fe/H] $<$
+0.67 (lower left to upper right).  
\label{grids}}
\end{center}}

The stellar population models can be used to 
disentangle age and metallicity, albeit in a model
dependent way, and derive {\it relative} ages and metallicities
for the NGC 1399 GCs. In particular we use the 
hydrogen Balmer lines which are strongly
age sensitive combined with magnesium and iron 
which  are largely metal sensitive. 

In Fig.~2 we show our GC data compared 
to the stellar population models of Maraston \& Thomas (2000)
extended to younger ages (Maraston 2001, private
comm.). 
To complement Mg$_2$, our most robust measure of metallicity, 
we use [MgFe] ([Mg $b$ $\times$ $\langle$Fe$\rangle$]$^{0.5}$), 
which minimizes the effects of the overabundance seen in
Mg $b$, whilst increasing measurement accuracy (Gonzalez 1993). 
We have chosen to derive ages and metallicities for 
the GCs using the Maraston grids, 
since these are calibrated on Galactic GCs. 
In Table 1, we list the derived ages and metallicities
of the GCs, obtained from the mean of the values
predicted in the upper two panels of Fig.~2. 
Uncertainties are obtained by perturbing the line-strengths by
their errors and re-deriving their ages and metallicities. 
We emphasize that these uncertainties represent the random 
measurement errors, and do 
not include possible systematic errors in the models themselves.

\vbox{
\begin{center}
\leavevmode
\hbox{%
\epsfxsize=9.2cm
\epsffile{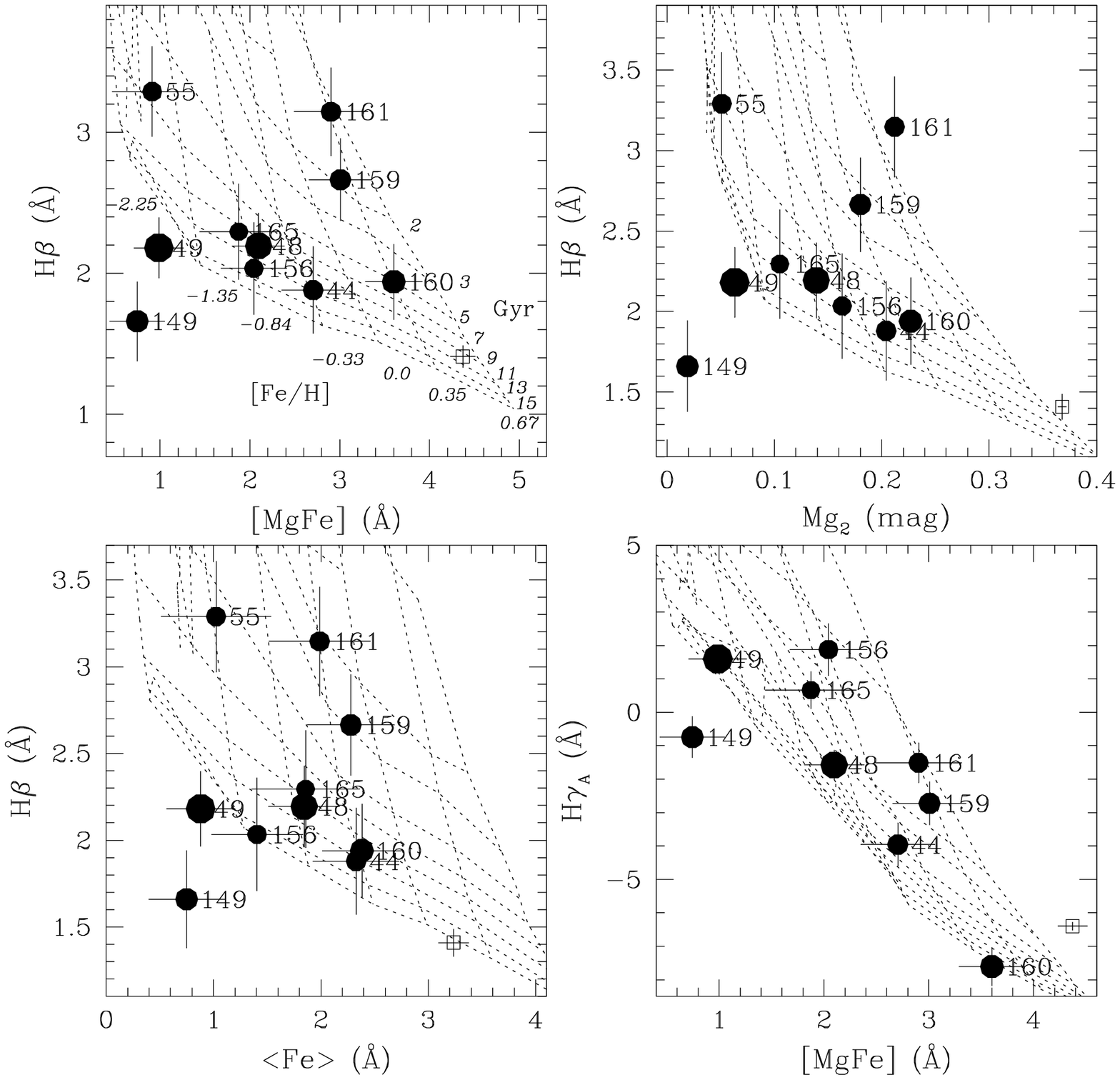}}
\figcaption{\small
Index measurements of NGC~1399 GCs compared
to the stellar population models of Maraston \& Thomas (2000).
The open square indicates the position of NGC~1399, 
from Kuntschner (2000). For the final ages and metallicities we
use the upper two grids. 
{\it Top left:} H$\beta$--[MgFe] grid. 
{\it Top right:} H$\beta$--Mg$_2$ grid. 
{\it Bottom left:} H$\beta$--$\langle$Fe$\rangle$ grid. 
{\it Bottom right:} H$\gamma_{\rm A}$--[MgFe] grid. The spectrum of GC 
\#55 does not include H$\gamma_{\rm A}$. 
\label{grids1}}
\end{center}}

In Fig.~3 we show integrated spectra of three GCs associated
with NGC~1399. As described below, two of the GCs (\#161, \#159) 
exhibit Balmer and metal line-stengths consistent with very young ages
(notice in particular the strong H$\beta$ and Mg features in the 
spectrum of \#161).
Cluster \#49 represents our highest S/N spectrum (H$\beta$ error
is $\pm$ 0.22~\AA) and is an example of an old 
metal--poor GC. 

The majority of the GCs are old (as found by Kissler-Patig \etal 1998) 
and are consistent with the 11 Gyr isochrone, within 2$\sigma$ of
their individual measurement errors. 
This age is consistent with the luminosity weighted
age for the central stellar population of NGC~1399, i.e. 
10 $\pm$ 2 Gyr using the line index measurement of Kuntschner
(2000). 
One metal-poor GC (
\#149) falls $below$ the oldest model isochrones, possibly 
reflecting the uncertainties in modelling horizontal-branches 
in the models. 

Interestingly, two of the GCs (\#159 and \#161) 
have very young inferred ages of $\sim$2 Gyrs. Such young ages
are consistent in all four model grids of Fig.~2, including 
H$\gamma_{\rm A}$. Worthey (1994) models also indicate young ages. 
Significantly, 
the age estimates from H$\beta$--[MgFe] and 
H$\beta$--$\langle$Fe$\rangle$  
are both consistent indicating that the 
non-solar [$\alpha$/Fe] ratios of the GCs are $not$ responsible.
It is important however to recognise that
the population models are somewhat uncertain
at very young ages, because they 
use fitting-functions derived only from 
old stars.

\vbox{
\begin{center}
\leavevmode
\hbox{%
\epsfxsize=8.0cm
\epsffile{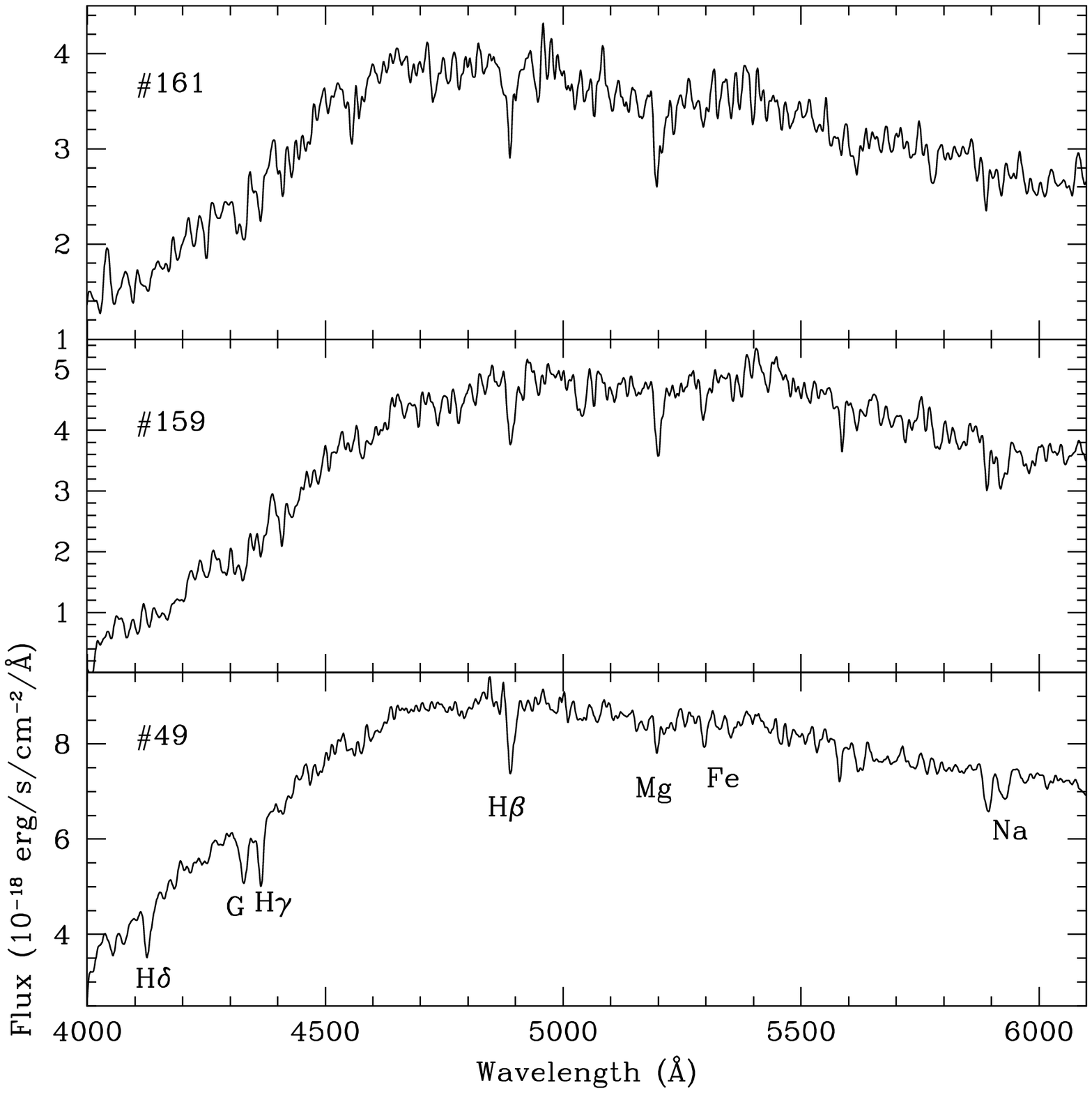}}
\figcaption{\small
Keck spectra of three globular clusters around NGC 1399. 
The spectra have been convolved with a wavelength-dependent 
Gaussian kernel to match the Lick/IDS resolution (8--11 \AA).
Clusters \#159, 161 are inferred to be metal-rich and young 
whereas \#49 is an example of a metal-poor old 
cluster. 
\label{spectra}}
\end{center}}

Kissler-Patig \etal (1998) also identified two GCs (\#83 and
\#102 from Grillmair 1992) with 
strong Balmer and metal lines, and potentially very young ages
albeit with large errors.  
One possibility raised by these authors was that their 
enhanced Balmer lines were due to the presence of 
blue horizontal branches (BHBs). 

de Freitas Pacheco \& Barbuy (1995), Maraston \& Thomas (2000) and Lee
\etal (2000) have investigated the potential influence of a BHB
on the H$\beta$ line index. According to Lee \etal a warm BHB can
raise the H$\beta$ EW in metal-rich GCs by up to 0.75~\AA, 
without significantly affecting the metallicity, {\it if} those GCs are
$\sim$ 4 Gyrs older than Galactic ones.  
For metal-rich Galactic GCs, with estimated ages $\sim$ 13 Gyrs, the
increase in H$\beta$ from a BHB is minimal. Thus assuming that
GCs \#159 and \#161 are truely metal-rich with [Fe/H] $\ge$ 0, 
then they must be $\ge$ 4 Gyr older than typical Galactic GCs 
in order for BHBs to explain their H$\beta$ EW. If we accept
current estimates for the oldest Galactic GCs to be 12.5
Gyrs, then 16.5 Gyr GCs would be incompatible with the current
best estimates for the age
of the Universe (Gnedin, Lahav \& Rees 2001).  
We note that future far-UV photometry may help to distinguish
between the presence of a BHB and a young age (Lee 2001, private
comm.)

So although we can not conclusively choose between these two
alternatives, both have interesting implications.
If the BHB interpretation is correct, it would be the first
detection of BHBs in metal-rich GCs of an elliptical galaxy
since the initial discovery in two metal-rich Galactic GCs by
Rich \etal (1997). It also implies that at least some 
GCs in NGC 1399 are
systemically older than their Galactic counterparts by at least 4
Gyrs, and hence in conflict with the age of the Universe under
certain cosmologies. 
The alternative is that these 
GCs, which are both metal-rich and have 
super-solar [Mg/Fe] abundance ratios,  
formed only $\sim$ 2 Gyrs ago. Whether these
GCs formed in an accreted satellite or {\it in situ} is not
clear, but to attain the enhanced abundance ratios would require
that they formed very soon after the first type II SNe. 
According to Thomas, Greggio \& Bender 
(1999), a recent merger would require
an extremely flat IMF to reproduce the observed $\alpha$ enhancement. 
However late epoch mergers of spiral disks, like the Milky Way, 
would {\it not} be expected to form $\alpha$ enhanced metal-rich
GCs (Goudfrooij \etal 2001). 

\section{Concluding Remarks}

From high S/N spectra and the stellar population
models of Maraston \& Thomas (2000) we find that the majority of globular 
clusters in NGC~1399 are old, similar to the luminosity-weighted
age of NGC~1399 itself. 
At least two clusters have super-solar
abundance ratios, again like the host galaxy.  
A super-solar abundance ratio for metal-rich stellar 
populations is a natural
outcome from a fast, clumpy collapse but may also be produced
by mergers if star formation has a sufficiently flat IMF (Thomas,
Greggio \& Bender 1999).
Two metal-rich GCs are reported with unusually
high H$\beta$ line strengths. It remains unclear whether this is
due to a young ($\sim$ 2 Gyr) age, or extremely old ($>$ 15
Gyr) age with a blue horizontal branch. However a conflict with
current cosmological parameters is avoided if the young age is
favoured. 

\section{Acknowledgments}\label{sec_ack}

We thank T. Bridges for the cluster photometry, 
A. Phillips for the use of his software, L. Schroder for help
with initial data reduction, and C. Maraston for providing
her model grids ahead of publication. We also thank S. Larsen, 
B. Gibson and C. Maraston for useful discussions. 
Part of this research was funded by NSF grant AST 9900732 and an ARC
grant. MB thanks the Royal Society. 
The data presented herein were obtained at the
W.M. Keck Observatory, which is operated jointly by
the California Institute of Technology and 
the University of California.

\begin{table}
\begin{scriptsize}
\begin{center}
\renewcommand{\arraystretch}{1.5}
\begin{tabular}{lcccccccccc}
\multicolumn{11}{c}{\scriptsize TABLE 1. CANDIDATE GLOBULAR CLUSTERS
AROUND NGC 1399}\\
\hline
\hline
ID$^a$ & H$\gamma$ & H$\beta$ & Mg~$b$ & Mg$_2$ & $<$Fe$>$ 
& T$_1$    & C--T$_1$ &
[Fe/H]$^{b}$ & Age$^{b}$ & V$_{helio}$ \\
   & (\AA) & (\AA) & (\AA) & (mag) & (\AA) 
& (mag) & (mag) & (dex) & (Gyr) & (km/s)\\
\hline
44 & -3.96$\pm$0.65 & 1.88$\pm$0.31 & 3.14$\pm$0.32 & 
0.204$\pm$0.008 & 2.33$\pm$0.40
& 21.19$\pm$0.01 & 1.75$\pm$0.02 & 
-0.3$\pm$0.3 & 10$^{+6}_{-5}$ & 1127$\pm$41\\
48 & -1.57$\pm$0.45 & 2.19$\pm$0.24  & 2.38$\pm$0.25 & 
0.139$\pm$0.006 & 1.85$\pm$0.33
& 20.58$\pm$0.01 & 1.43$\pm$0.01 & 
-0.8$\pm$0.2 & 10$^{+4}_{-4}$ & 1831$\pm$48\\
49 & 1.60$\pm$0.37 & 2.18$\pm$0.22  & 1.11$\pm$0.24 & 
0.063$\pm$0.006 & 0.88$\pm$0.32
& 20.48$\pm$0.01 & 1.22$\pm$0.01 & 
-1.7$\pm$0.2 & $\sim$ 15 & 1618$\pm$64\\
55 & ...& 3.29$\pm$0.32 & 0.81$\pm$0.39 & 0.051$\pm$0.009 & 
1.03$\pm$0.51
& 20.87$\pm$0.01 & 1.14$\pm$0.01 & 
-1.9$\pm$0.3 & 7$^{+3}_{-3}$ & 1364$\pm$65\\
149 & -0.75$\pm$0.62 & 1.66$\pm$0.28  & 0.74$\pm$0.27 & 
0.019$\pm$0.007  & 0.75$\pm$0.35
& 20.84$\pm$0.01 & 1.14$\pm$0.01 &
-2.2$\pm$0.3 & $\geq$ 15 & 1361$\pm$107\\
156 & 1.88$\pm$0.78 & 2.03$\pm$0.33  & 2.97$\pm$0.32 & 
0.163$\pm$0.008 & 1.41$\pm$0.42
& 21.12$\pm$0.01 & 1.50$\pm$0.02 & 
-0.7$\pm$0.4 & 11$^{+7}_{-7}$ & 1662$\pm$43\\
159 & -2.73$\pm$0.65 & 2.66$\pm$0.29  & 3.97$\pm$0.30 & 
0.180$\pm$0.008 & 2.28$\pm$0.41
& 21.08$\pm$0.01 & 1.47$\pm$0.02 &
0.1$\pm$0.3 & 2.3$^{+1}_{-1}$ & 1579$\pm$41\\
160 & -7.61$\pm$0.58 & 1.94$\pm$0.27 & 5.45$\pm$0.27 & 
0.227$\pm$0.007 & 2.38$\pm$0.37 & 20.75$\pm$0.01 &
1.72$\pm$0.01 & 
0.2$\pm$0.3 & 7$^{+6}_{-4}$ & 1378$\pm$32\\
161 & -1.51$\pm$0.59 & 3.15$\pm$0.31  & 4.23$\pm$0.36 & 
0.212$\pm$0.009 & 1.99$\pm$0.47
& 20.89$\pm$0.01 & 1.45$\pm$0.01 &
0.3$\pm$0.3 & 1.6$^{+1}_{-1}$ & 1506$\pm$45\\
165 & 0.67$\pm$0.56 & 2.30$\pm$0.34 & 1.89$\pm$0.38 & 
0.105$\pm$0.009 & 1.86$\pm$0.50
& 20.82$\pm$0.01 & 1.43$\pm$0.01 & 
-1.0$\pm$0.3 & 11$^{+7}_{-5}$ & 2020$\pm$38\\
\hline
\end{tabular}
\end{center}
\tablenotetext{a}{\scriptsize
globular cluster ID number from Grillmair (1992).
}
\tablenotetext{b}{\scriptsize
metallicity and age are derived using the single stellar population
models of Maraston \& Thomas (2000). 
}
\end{scriptsize}
\end{table}

\end{document}